\documentclass[12pt]{article}
\usepackage{amsmath, amssymb}
\usepackage{graphicx}
\usepackage[top=2.5cm, left=3cm, bottom=3cm, right=2cm]{geometry}
\author{Azizollah Azizi\thanks{aziz@shirazu.ac.ir} 
	\and 
	Shaghayegh Parkami\thanks{parviz7207@gmail.com}\\
	Department of Physics, Shiraz Univesity, Shiraz, I.R.Iran}
\title{Coupled Sine-Gordon and $\phi^4$ System}
\newcommand{\olsi}[1]{\,\overline{\!{#1}}} 
\begin{document}
	\baselineskip 16.7pt
\maketitle
\begin{abstract}
Coupling the fields may lead to the emergence of new phenomena. In the realm of classical fields and nonlinear systems, extensive research has been conducted on their solitary and soliton solutions. In the conducted studies, typically two $\phi^4$ systems, or two sine-Gordon systems, have been coupled. The sine-Gordon system exhibits diverse solutions, all well-behaved, with its soliton solutions fully understood. On the other hand, the $\phi^4$ system, which is significant in field theory, has solitary solutions, but these solutions are not solitonic. For example, from a pair of kink and antikink, we cannot construct a bound state; or that after a collision, these two solutions do not revert to their initial status and become disrupted. In this study, we couple a $\phi^4$ system with a sine-Gordon system to impart stability from the sine-Gordon system to the $\phi^4$ system. We have demonstrated that for a coupled $\phi^4$ and sine-Gordon system, this expectation is somewhat met.
\end{abstract}
\section{Introduction}
In physics, the study of nonlinear fields is very important in both non-relativistic and relativistic districts. In the relativistic fields and in two dimensions (1+1), two sine-Gordon and $\phi^4$ systems have been studied more than the others. The sine-Gordon equation appears in numerous and unrelated areas \cite{ref1}. This system with the simplified Lagrangian
\begin{equation}\label{sg_lag}
	\mathcal{L}=\frac{1}{2}\partial_\mu\partial^\mu\psi(x,t)-\left(1-\cos{\psi}(x,t)\right),
\end{equation}
has many well-known soliton solutions. In all cases, the study of the interaction between each two solutions confirms the stability of these solutions. Among the interesting cases is the formation of a bound state consisting of two kink and antikink solutions of this system (with opposite topological charges). For this purpose, it is enough to place two solutions without relative velocity (or with low relative velocity) and at a short distance from each other, so that we can see their periodic motion (without destruction or loss of energy). In addition, if two solutions collide with a significant relative velocity, we will witness an elastic collision.
\par
And, the $\phi^4$ field in the field theories and statistical mechanics, especially in the topics related to the spontaneous symmetry breaking, has very important role, both in four (1+3) and in two dimensions (1+1). Solitary solutions of this system with the simplified Lagrangian
\begin{equation}\label{phi4_lag}
	\mathcal{L}=\frac{1}{2}\partial_\mu\partial^\mu\phi(x,t)-\frac{1}{4}\left(1-\phi^2(x,t)\right)^2
\end{equation}
in (1+1) dimensions are investigated. Unlike the sine-Gordon system, a pair of a kink and an antikink of this system does not form a bound state, and the two solutions after fusion collapse and destroy. The study of collision of these two solutions for high relative velocities also shows the lack of stability of the solutions after the separation of the two solutions from each other. As a result, the $\phi^4$ system solutions are evaluated as a system with solitary solutions but without soliton solutions.
\par
In the field theories, fields are coupled (mixed) for various purposes such as interaction, renormalization, mass production, symmetry breaking, and more. In the theory of classical relativistic fields in (1+1) dimension, this is very common. Investigating the characteristics of the solutions of such a system of coupled fields, especially the investigation of solitary and soliton solutions, is very interesting.
\par
In literature, researchers usually present a potential that either includes two $\phi^4$ fields \cite{ref2,ref3,ref4,ref5} or includes two sine-Gordon fields \cite{ref6,ref7}. In most of these works, the existence of solutions and some of their characteristics have been studied; or stationary solutions have been obtained by using different algorithms.
\par
In this paper, we have done two special things. First, we have coupled a $\phi^4$ field to a sine-Gordon field. What do we mean by this modulation? We module a ``loose'' system with no soliton solution, to a ``strong'' system with stable solutions, perhaps the stronger system shares its properties with the weaker system. Second, we have dealt with the collision of the obtained stationary solutions of the system. We have studied our expectations from the modulation of the two systems.
\section{The coupled system of sine-Gordon and $\phi^4$ fields}
\begin{figure}
	\centering
	\includegraphics[height=0.4\textheight]{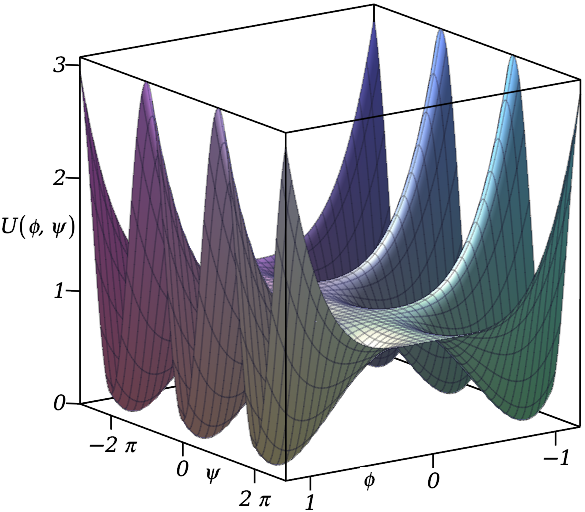}
	\caption{The potential of coupled sine-Gordon and $\phi^4$ fields, and its minimum points. \label{fig1}}
\end{figure}

To pair two fields $\phi^4$ and sine-Gordon, there are many choices. We introduce the potential with some considerations. We consider the potential to be non-negative, so that at vacuum points, the energy is minimal and equal to zero. We consider the vacuums of this system as $\left\{\phi=\pm1,\ \psi=2n\pi,\  n\in\mathbb{Z}\right\}$, which are $\phi^4$ and sine-Gordon systems vacuums. The members of this set (in pairs) must satisfy the equations of motion and minimize the energy. In addition, since the focus of this study is on the $\phi^4$ system, we import its potential $(\phi^2 - 1)^2$ intact, and to couple two systems, we multiply the sine-Gordon system potential, $(1 - \cos{\psi})$, by a factor $\phi^2$. So we introduce the following potential for our coupled system
\begin{equation}\label{pot}
	U\left(\phi,\psi\right) = \frac{1}{4}\left({1-\phi}^2\right)^2+g\,\phi^2\left(1-\cos{\psi}\ \right).
\end{equation}
In the above relation, $g$ is the coupling coefficient of the two fields, which we consider it to be equal to 1 in this article. This potential and the location of its minima are shown in Figure \ref{fig1}. The Lagrangian of this coupled system in two dimensions is
\begin{equation}\label{copsys}
	\mathcal{L}=\frac{1}{2}\partial_\mu\phi\left(x,t\right)\partial^\mu\phi\left(x,t\right)+\frac{1}{2}\partial_\mu\psi\left(x,t\right)\partial^\mu\psi\left(x,t\right)-U\left(\phi,\psi\right),
\end{equation}
and the equations of motion are
\begin{align}
	\square\phi &= -\frac{\partial  U}{\partial\phi} = -2\phi\left(1-\cos{\psi}\right)-\phi\left(\phi^2-1\right), \label{phidyn}\\
	\square\psi &=-\frac{\partial U}{\partial\psi} = -\phi^2\sin{\psi.} \label{psidyn}
\end{align}
The energy density, $\mathcal{E}$, and the total energy, $E$, for this system are defined as below
\begin{align}
	\mathcal{E}&=T_0^0=\frac{1}{2}\left(\frac{\partial\phi}{\partial\ x}\right)^2+\frac{1}{2}\left(\frac{\partial\phi}{\partial t}\right)^2+\frac{1}{2}\left(\frac{\partial\psi}{\partial\ x}\right)^2+\frac{1}{2}\left(\frac{\partial\psi}{\partial t}\right)^2+U\left(\phi,\psi\right),\label{energy_den}\\
	E&=\int_{-\infty}^{\infty}{\mathcal{E}\ dx}\label{energy}.
\end{align}
Two topological charges for this system are introduced as follows
\begin{align}
	Q_\phi&=[\phi\left(+\infty\right)-\phi\left(-\infty\right)]/2,\label{phi_charge}\\
	Q_\psi&=[\psi(+\infty)-\psi\left(-\infty\right)]/2\pi. \label{psi_charge}
\end{align}
The charge $Q_\phi$ for kink (antikink) is $+1$ ($-1$), and zero for the mass; and $Q_\psi$ for kink (antikink) is $+1$ ($-1$), or in general an integer number.
\section{Investigating the solutions}
To obtain stationary solutions (independent of time) in our numerical algorithm, we minimize the energy of relation (\ref{energy}). For this purpose, we fix the boundary conditions (asymptotic behavior of the solution), and start with suitable test functions and achieve the desired solution (more details are in the appendix). 
\par
The solutions of this coupled system can be divided into four categories:
\begin{enumerate}
	\item
In the first category, $\phi$ goes from a vacuum at $x=-\infty$ to adjacent vacuum at $x=+\infty$, (for example, from $\phi=-1$ to $\phi=+1$), while $\psi$ goes from a vacuum to the same one (for example, from zero to zero, or from $2\pi$ to $2\pi$). We call this category horizontal, and denote it by H. For this category $Q_\phi=\pm 1$ and $Q_\psi=0$.
	\item 
In the second category, unlike the first category, $\psi$ goes from a system vacuum to an adjacent one, while $\phi$ goes from one vacuum to the same one. We call this category vertical, and denote it by V. For this category $Q_\phi=0$ for and $Q_\psi=n,\ n\in\mathbb{N}$
	\item 
The third category are solutions that both $\phi$ and $\psi$ fields change their vacuums from $-\infty$ to $+\infty$. We call this category diagonal, and denote it by D. For this category $Q_\phi=\pm1$ and $Q_\psi=n,\ n\in\mathbb{N}$.
	\item 
In the last item, none of the two fields change their vacuums at the both ends. So, for this category $Q_\phi$ and $Q_\psi$ are zero.
\end{enumerate}
\par
\begin{figure}[t!]
	\centering
	\includegraphics[height=0.37\textheight]{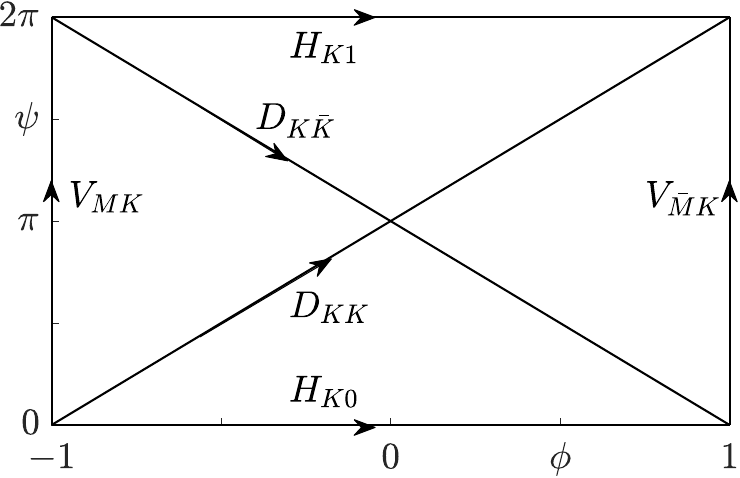}
	\caption{System solutions for vacuums $\left\{\phi=\pm1,\ \psi=\pm2\pi\right\}$. \label{fig2}}
\end{figure}

Figure \ref{fig2} shows these descriptions schematically.
\par
It is expected that the shape of the solution from one vacuum to the same vacuum, is a bell-shaped curve, which we denote it by $M$ (mass). It is also expected that the solution from one vacuum to an adjacent vacuum is in the form of a kink, which we denote it by $K$. As usual, every solution has an anti-solution. The point is that, the solution $\phi:-1\mapsto-1$, is a mass in which the convexity of the shape is upward, and we denote it by $M$. We show instead, for the solution that $\phi:+1\mapsto+1$, the solution is convex downward, which we denote it by $\olsi{M}$. By the way, from $-1$ to $-1$ there exists only $M$, and from $+1$ to $+1$ only $\olsi{M}$.
\par
For horizontal solutions (H), which we call it $Kn, n\in\mathbb{N}$, the coupled system is reduced to a $\phi^4$ single system; That is, the solutions are the known kink and antikink solutions of the $\phi^4$ system, and $\psi=(2\pi)n,\ n\in\mathbb{Z}$ are the obvious vacuum solutions of the sine-Gordon system.
\par
\begin{figure}[t]
	\centering
	\includegraphics[height=0.34\textheight]{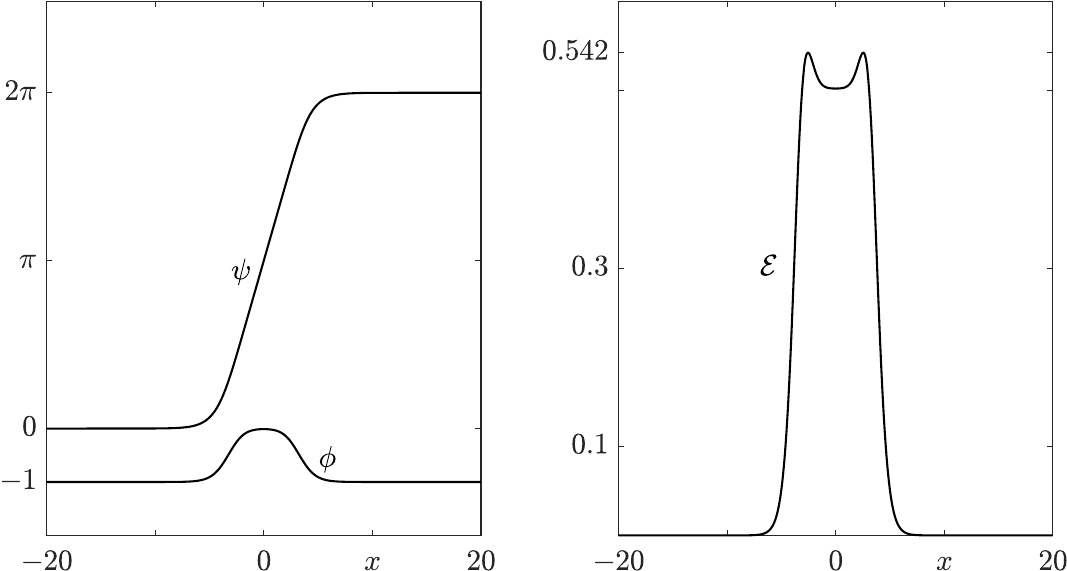}
	\caption{The left side are the fields, and the right side are the energy density for a vertical solution $MK$. \label{fig3}}
\end{figure}

Instead, for the vertical case (V), there is non-obvious solution. In this solution, the curve corresponds to $\phi$ is a mass, and the solution corresponding to the sine-Gordon is a kink (Figure \ref{fig3}). It can be seen that in a significant interval around the symmetry axis, i.e $x=0$, the slope of the curves remain constant.
\par
The interesting thing that is observed, is the twin-peaked form of the energy density. 
\begin{figure}[!b]
	\centering
	\includegraphics[height=0.35\textheight]{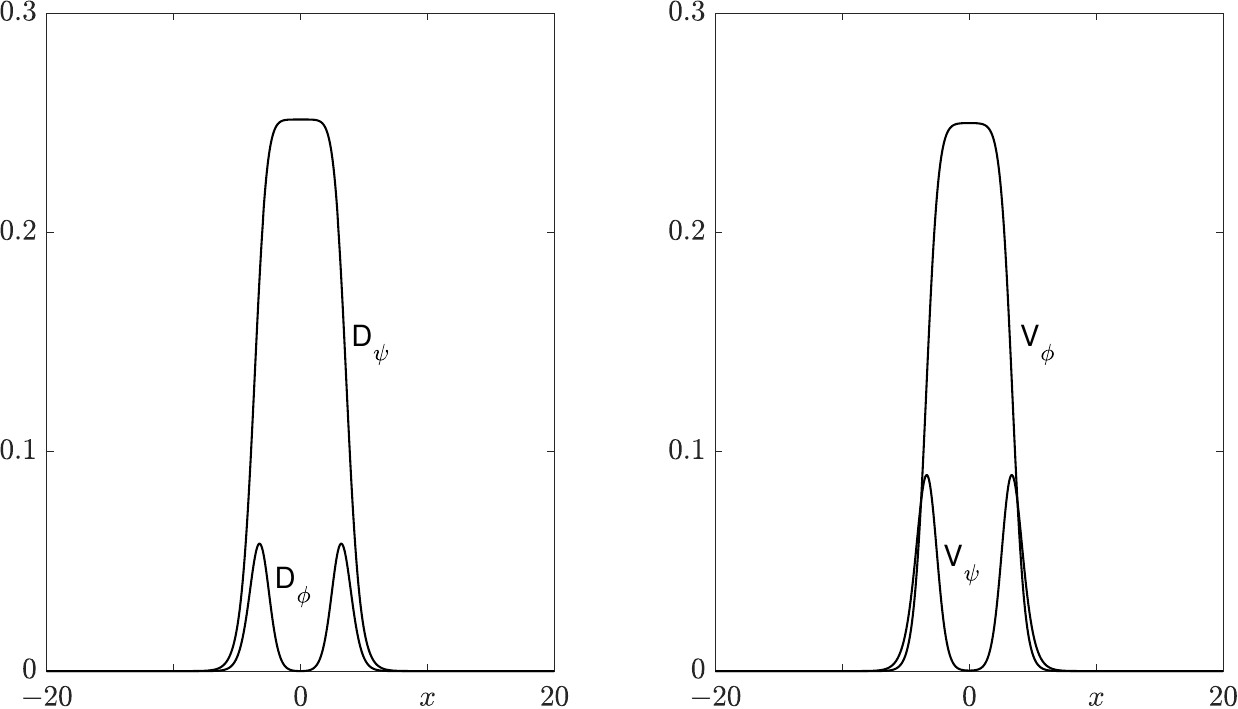}
	\caption{For a $MK$ solution: Left side, $D_\phi=\left(\frac{1}{2}\right)\left(\frac{d\phi}{dx}\right)^2$ and $D_\psi = \left(\frac{1}{2}\right)\left(\frac{d\psi}{dx}\right)^2$, and right side, $V_\phi = \frac{1}{4}\left(1-\phi^2\right)^2$, $V_\psi=\phi^2(1-\cos{\psi)}$. \label{fig4}}
\end{figure}
In Figure \ref{fig4}, the components of the energy density, the square of the first derivative of two functions $\phi(x)$ and $\psi(x)$ and two parts of the potential energy are plotted, to clarify the origin of the twin peaks of the energy density. Although $\phi=0$ is not a system vacuum, it locally acts like a vacuum (note the flatness of $\phi$ around $x=0$ in Figure 3 left).
Mass (rest energy) of the solution $MK$ is obtained as 4.1994, which is less than the average mass of $\phi^4$ ($2\sqrt{2}/3=0.9428$), and Sine-Gordon kink (8).
\par
\begin{figure}
	\centering
	\includegraphics[height=0.34\textheight]{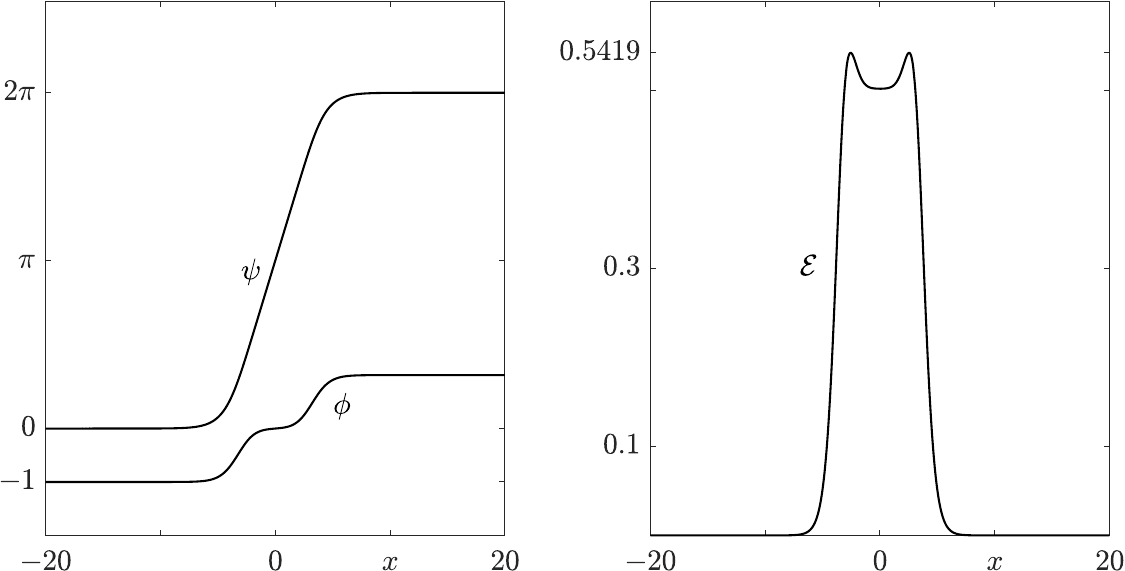}
	\caption{Fields and energy density for Diagonal solution $KK$. Kink of $\phi$ has a step. \label{fig5}}
\end{figure}

For the diagonal case, for example $KK$, the shape of both fields are kinks, with an explanation that the kink of the field $\phi$ has a step (Figure \ref{fig5}). The mass of this solution is 4.1996, with a slight difference with the solution of $MK$ (within the numerical calculation error). The step nature of $\phi$ is another interesting property of a solution in this system.
\par
By comparing the solutions $MK$ and $KK$, it has been observed that the kink of the sine-Gordon field in the two solutions are almost identical. In addition, the absolute value of the kink of $\phi$ in Figure 5 (which is an odd function), i.e. $|K|$, is almost coincides with the mass curve $M$ in Figure 3 (which is an even function). But in these two adaptations, there is a significant difference. In the last case case, for $|K|$ to completely coincides on $M$, they must have equal left and right derivatives of all orders, which is not possible for an ordinary odd functions (with Taylor expansion). As a result, the difference between these two solutions is worth considering.
\par
For the case that both $Q_\phi$ and $Q_\psi$ are zero, we expect the solution of both fields to be bell-shaped. Our numerical efforts in this case did not lead to a non-obvious (non-vacuum) solution, and there probably only obvious (vacuum) solutions exist for this case! States like mass for field $\phi$ and breather for the sine-Gordon field cannot be obtained by numerical algorithms.
\par
From the symmetries of Lagrangian and equations of motion under transformations $\phi\rightarrow\pm\phi$ and $\psi\rightarrow\pm\psi$, it is enough to find only one member of the two sets \{$KK$, $K\olsi{K}$, $\olsi{K}K$, $\olsi{K}\olsi{K}$\}, and \{$MK$, $\olsi{M}K$, $M\olsi{K}$, $\olsi{M}\olsi{K}$\}.
\par
The solutions we reviewed in this section are all solitary solutions. In the next section, we will study the collisions between these solutions.
\section{Stability of solutions}
Investigation of the collision between the solitary solutions led us to notice different phenomena, which we will discuss below.
\par
First, we investigate the collision between two V solutions. We come across an interesting phenomenon, and that is that there are two critical velocities. If the relative velocity of the two solutions is lower than the first critical velocity, $v_\mathrm{cr}^{(1)}=\mathrm{0.6139}$, or higher than the second critical velocity, $v_\mathrm{cr}^{(2)}=\mathrm{0.7487}$, the two solutions will separate from each other after merging (collision) without any problems, i.e. $MK+MK \rightarrow MK+MK$ (Figure 6, top row).
\begin{figure}
	\centering
	\includegraphics[width=.95\textwidth]{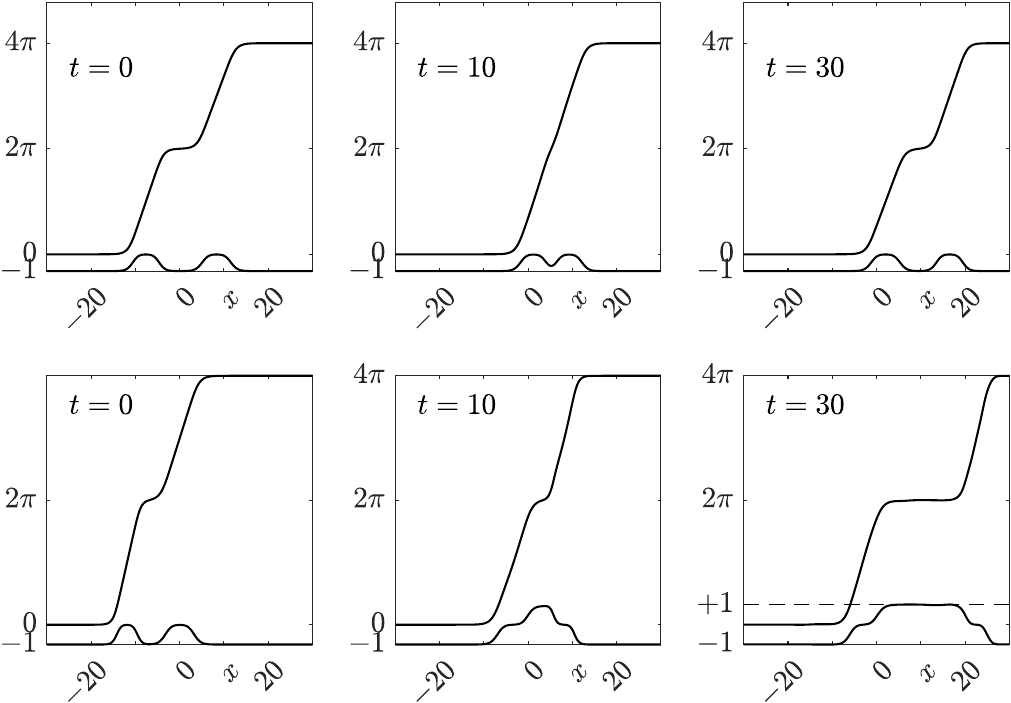}
	\caption{In the top row, two $MK$ solutions with a relative velocity of $0.4$; at $t=0$ started to collide (left), then merged together (middle) and later separated (right). In the bottom row, two $MK$ solutions at $t=0$ started collision with a relative velocity of $0.7$, (left), then mixed together (middle) and finally emerged as a pair $KK$ and $\olsi{K}K$ (right). \label{fig6}}
\end{figure}
If the relative velocity of the two solutions is between these two critical velocities, i.e. $v_\mathrm{cr}^{(1)}<v<v_\mathrm{cr}^{(2)}$, after intermingling of the two initial solutions, two other different solutions will be born, i.e. $MK+MK \rightarrow KK+\olsi{K}K$ (Figure 6, bottom).
\par
In the collision between feasible diagonal solutions, we also have a phenomenon similar to the former case, i.e. there are also two critical velocities. So, if the relative velocity of the two solutions is less than the smaller critical velocity, $v_\mathrm{cr}^{(1)}=\mathrm{0.6113}$, or more than the larger critical velocity, $v_\mathrm{cr}^{(2)}=\mathrm{0.7498}$, the two solutions will separate from each other without changing after the collision. And if $v_\mathrm{cr}^{(1)}<v<v_\mathrm{cr}^{(2)}$, after collision they convert to two solutions of the type V, that is, $KK+\olsi{K}K\rightarrow MK+MK$.
\par
Another case is to study the collision between V and D solutions. In this case, as in the previous two cases, there are two critical velocities; $v_\mathrm{cr}^{(1)} = \mathrm{0.6130}$ and $v_\mathrm{cr}^{(2)} = \mathrm{0.7505}$. For example, let $KK$ collides with $MK$. If the relative velocity is outside the range of critical velocities, the solutions themselves will be reappeared after the collision; and if the relative velocity is between the critical velocities, $KK$ and $\olsi{M}K$ will be achieved.
\begin{figure}[!t]
	\centering
	\includegraphics[width=.95\textwidth]{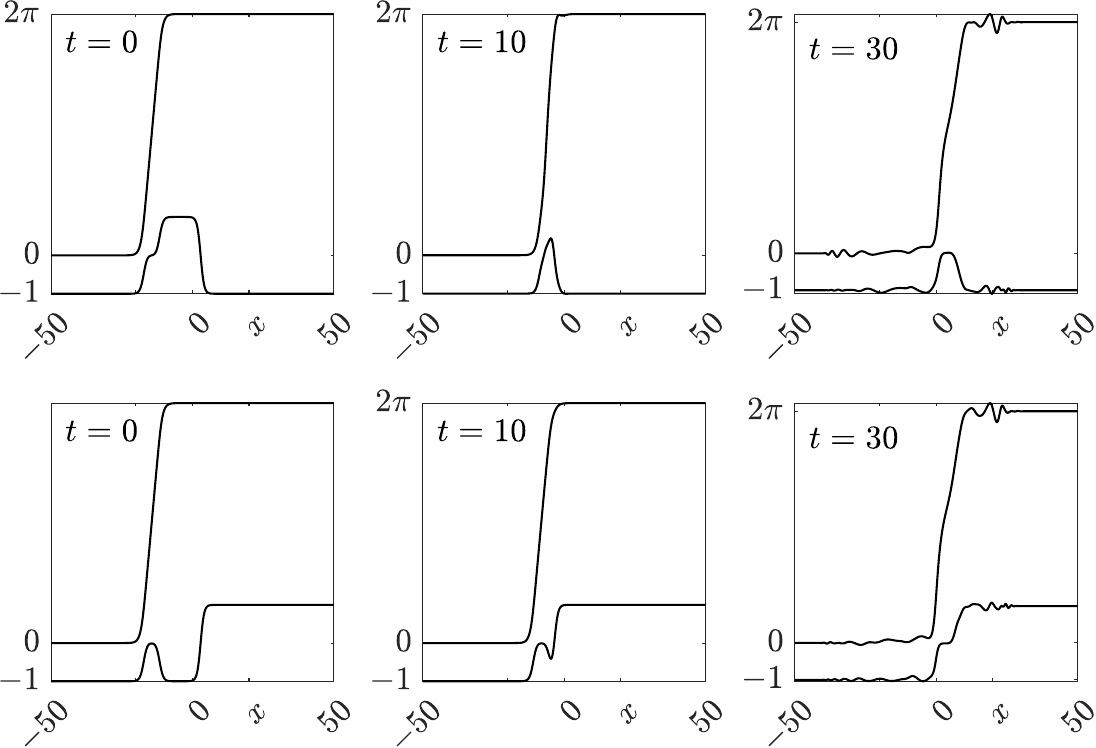}
	\caption{Top row $KK+\olsi{K}0\rightarrow MK+$``energy lost", and bottom row $MK+K0\rightarrow KK+$``energy lost''. \label{fig7}}
\end{figure}
\par
Finally, we note the collision of horizontal solutions H with the other solutions. The study of the interaction of these solutions with themselves is the study of the interaction of the $\phi^4$ system solutions, which we mentioned earlier.
In the collision of a horizontal solution $\olsi{K}0$ (``$\olsi{K}$'' is $\phi^4$ antikink, and ``$0$'' is a sine-Gordon vacuum), with a diagonal solution, for example $KK$, after the mixing of two solutions, with some energy loss, the diagonal solution $\olsi{M}K$ appears (Figure 7 top). 
If the horizontal solution $K0$ is collided with a vertical solution, for example $MK$, this time some energy is wasted, and a diagonal solution $KK$ is obtained (Figure 7, bottom). These results are valid for the whole range of velocities, and are understood by the principle of conservation of charges.
\par
In all collisions between solutions that a kink and an antikink of sine-Gordon are placed in front of each other, after the collision (mixing) the solutions are messed up and destroyed. For example if $M\olsi{K}$ collides with $MK$, after collision, the total energy is radiated and two straight lines (vacuums) are obtained. This actually confirms that a solution in which neither of the two fields change the vacuum (the total charge is zero) is an obvious solution.
\section{Conclusions}
The innovation of this paper is in coupling of a non-solitonic system $\phi^4$ to a solitonic system sine-Gordon. The solitary solutions of the system have been obtained using the energy minimization algorithm. All these solutions are stable and independent of the observer as long as they are free (i. e. they maintain their nature over time). In this between, two diagonal and vertical solutions are particularly important. These two solutions, which have different topological characteristics, seem to be two degenerate states. In our numerical calculations, there is a slight difference in the masses (and critical velocities), which is apparently within the error range of the numerical calculations.
\par
Numerical algorithms have a systematic error, which usually increases when the step size is reduced beyond a limit. Considering this matter and of course the speed of the computers, we could not increase the accuracy, and we preferred not to ignore the mentioned minor differences and report them.
\par
In all collisions between the solutions D and V, similar phenomena occur. There are two critical velocities. These velocities (up to a numerical error!) do not depend on the colliding objects. Collisions outside the range of critical velocities are perfectly elastic, and within the range of critical velocities we see a phase transition. The origin of this phase transition is still not clear to us.
\par
In the initial design of the coupled system, due to the persistence of soliton solutions of the sine-Gordon system in collisions, we somehow expected a bound state, but the result was not as expected. For example, in the collision
$$MK\left(0\rightarrow2\pi\right)+M\olsi{K}(2\pi\rightarrow0),$$
the charge of each of the two masses is zero (that is, we do not expect any attraction or repulsion from this part), but the opposite charge of the kink and anti-kink of $\psi$ could have caused a bound state, which did not happen! What happens is that after mixing the fields, the solutions get mixed up, and after all the energy is wasted, the obvious solution $\phi=-1,\ \psi=0$ is obtained.
\par
In addition to the system vacuums, $\phi=0$ and $\phi=$ ``constant" also satisfy the equations of motion (\ref{phidyn}) and (\ref{psidyn}), and is therefore a solution of the system. According to the potential (\ref{pot}) and Figure \ref{fig1}, this solution indicates a local vacuum of the system, and the potential at this solution is not an absolute minimum, but a local maximum.
\par
The unique similarity between $D_\psi$ with $V_\phi$, and $D_\phi$ with $V_\psi$ in Figure (4) can be an inspiration for examining and analyzing the system's solution.
\par
We introduced the coupling constant $g$ in the potential (\ref{pot}), but we did all the calculations of this paper with $g=1$. By changing $g$, the strength effect of field $\psi$ on the system, one may investigate its effect on the results. For example, it is possible to control the step shape of the $\phi$ kink in Figure 5, or the twin-peak form of the energy density in Figure 3. Additionally, its effect on collisions, critical velocity(ies) and post-collision outcomes can be investigated. The results of this research will be presented in future works.
\section*{Appendix: Description of numerical methods}
To obtain the stationary solutions of the coupled system (\ref{copsys}), we work in a framework where the solutions are at rest, so the equations of motion (\ref{phidyn}) and (\ref{psidyn}) become as below
\begin{align}
	&{{{\frac{d^2\phi}{dx^2}}}}-2\phi\left(1-\cos{\psi}\right)-\phi\left(\phi^2-1\right)=0,\label{phistat}\\
	&{\frac{d^2\psi}{dx^2}}-\phi^2\sin{\psi}=0.\label{psistat}
\end{align}
Instead of solving coupled differential equations (\ref{phistat}) And (\ref{psistat}) numerically (which is a difficult job), by minimizing the energy
\begin{equation}\label{energy_stat}
	E=\int_{-\infty}^{\infty}\left\{\frac{1}{2}\left(\frac{\partial\phi}{\partial\ x}\right)^2+\frac{1}{2}\left(\frac{\partial\psi}{\partial\ x}\right)^2+U\left(\phi,\psi\right)\right\}dx,
\end{equation}
we get the solutions. It is clear that fields that minimize energy, satisfy the equations of motion.
\par
Our algorithm is getting the solution using an iterative method. We start with two optional functions for two fields $\phi$ and $\psi$, with appropriate asymptotic behavior (boundary conditions). For example, to obtain a vertical solution MK in which $\phi$ goes from $-1$ to $-1$ and $\psi$ from $0$ to $2\pi$, the initial functions $\phi_0=-1$ and $\psi_0=4\arctan{(\exp{(x))}}$ are good choices (the first is the trivial vacuum solution of the $\phi^4$ system, and the second is the kink solution of the sine-Gordon system). It is clear that these two functions do not satisfy equations of motion (\ref{phistat}) and (\ref{psistat}).
\par
We consider two loops. In the inner loop, the counter sweeps from the beginning to the end of the $x$ interval. In this loop, at each point $x_i$ the value of each of the two functions is increased or decreased independently by a very small increment, then the energy of the system is calculated and be compared with the energy of the system before the changing. Now we consider whichever one has less energy, and choose it as the solution at this step. Then we go to the next point to finally get out of the inner loop. The outer loop is the iteration loop. The condition for exiting this loop is that the energy difference in this iteration and the previous iteration become smaller than an error that we specified at the beginning. In this way, with an acceptable number of iterations, we get a suitable solution.
\par
We can check the correctness of the final solution by putting them in the equivalence difference equations of differential equations (\ref{phistat}) and (\ref{psistat}). Or, put the obtained solution as the initial condition (solution at time $t=0$) in the time-dependent equations (\ref{phidyn}) and (\ref{psidyn}). If this solution remains undamaged and without distortion, and stay at rest with the passing of time, it is a correct and a solitary solution of the system.
\par
In order to move the solutions, we must include the Lorentz boost. For example, if we call one of the two stationary functions $f(x)$, and want to move this solution at $t=0$ from the place $x_0$ with velocity $v$, we have to calculate $f(\gamma(x-vt-x_0))$, which $\gamma=(1-v^2)^{-1/2}$ is the Lorentz factor (the velocity of light is taken to $c=1$). The point to be noted here is that the function $f$ is not known at new points, and for this purpose we use interpolation. 

\begin{thebibliography}{10}
	%
	\bibitem{ref1} J. Cuevas-Maraver, P. G. Kevrekidis and F. Williams (Editors), ``The Sine-Gordon Model and its Applications'', Springer International Publishing, Switzerland, (2014).
	%
	\bibitem{ref2} M. Mohammadi and N. Riazi, ``The Affective Factors on the Uncertainty in the Collisions of the Soliton Solutions of the Double Field Sine-Gordon System'', Commun. Nonlinear Sci. Numer. Simul. \textbf{72}, 176-193 (2019).
	%
	\bibitem{ref3} N. Riazi and M. Peyravi, ``Families of stable and metastable solitons in coupled system of scalar fields'', Int. J. Mod. Phys. A, \textbf{27}, 1250006 (2012).
	%
	\bibitem{ref4} D. Bazeia, M. J. dos Santos and R. F. Ribeiro, ``Solitons in systems of coupled scalar fields'', Phys. Lett. A, \textbf{208}, 84-88 (1995).
	%
	\bibitem{ref5} A. Alonso-Izquierdoa, D. Miguelez-Caballero and L. M. Nieto, ``Wobbling kinks and shape mode interactions in a coupled two-component $\phi^4$ theory'', Chaos, Solitons \& Fractals, \textbf{178}, 114373 (2024).
	%
	\bibitem{ref6} N. Riazi, A. Azizi, and S. M. Zebarjad, ``Soliton Decay in Coupled System of Scalar Fields'', Phys. Rev. D, \textbf{66}, 065003 (2002).
	%
	\bibitem{ref7} D. K. Campbell, M. Peyrard and P. Sodano,``Kink-Antikink Interactions in the Double {Sine-Gordon} Equation'', Physica D, \textbf{19}, 165-205 (1986).
	%
\end{thebibliography}
\end{document}